\documentclass{article}
\usepackage{amsfonts}
\usepackage{amsmath}

\newtheorem{theorem}{Theorem}

\newtheorem{example}[theorem]{Example}

\newtheorem{proposition}[theorem]{Proposition}
\newtheorem{remark}[theorem]{Remark}

\input{tcilatex}

\begin{document}

\title{Equations of electromagnetism in some special anisotropic spaces}
\author{Nicoleta BRINZEI \\
Transilvania University, Brasov, Romania \and Sergey SIPAROV \\
Academy of Civil Aviation, St. Petersburg, Russia}
\maketitle

\begin{abstract}
We show that anisotropy of the space naturally leads to new terms in the
expression of Lorentz force, as well as in the expressions of currents.
\end{abstract}

\section{Introduction}

Studying anisotropic spaces has an obvious meaning with regard to physical
interpretations. The direction dependence of the metric could cause the
appearance of motion dependent forces \cite{Newt-7} associated with inertial
forces in the accelerated frames. In case there is a physical vector field -
an electromagnetic one - in the anisotropic space, this may lead to the
appearance of the extra Lorentz type forces or extra currents that could
reveal themselves in a special laboratory environment or even in Nature.
From mathematical point of view, it is possible to treat the problem in the
purely Finslerian setting when $g_{ij}=\dfrac{1}{2}\dfrac{\partial ^{2}%
\mathcal{F}^{2}}{\partial y^{i}\partial y^{j}}$ for some 2-homogeneous in $y$
function $\mathcal{F}=\mathcal{F}(x,y)$ or introduce a more general type of
anisotropic metric that could explicitly give extra terms in the equations
of geodesics.

If we take into account the $y$-dependence of the fundamental metric tensor
in anisotropic spaces, then the components of an electromagnetic-type tensor 
$F_{ij},$\ $F_{~j}^{i},$\ $F^{ij}$\ could depend on the directional
variables. In order to make sure of this, notice the following. In \textbf{%
isotropic} (pseudo-Riemannian) spaces with $R_{ij}=0$, the components of the
free electromagnetic potential 4-vector $A^{i}=A^{i}(x)$ obey de Rham
equations: 
\begin{equation*}
A_{~~;\nu }^{\mu ;\nu }=0~\ 
\end{equation*}
that is, 
\begin{equation*}
g^{\nu \rho }(x)\nabla _{\nu }\nabla _{\rho }(A^{\mu })=0.
\end{equation*}

When passing to \textbf{anisotropic} spaces with metric $g_{ij}=g_{ij}(x,y),$
the solution of such an equation would generally depend on directional
variables (not to mention that the equation itself could become more
complicated). So, it is meaningful to take into consideration the case when
the potential 4-vector depends on the directional variables $y=(y^{i})$, 
\begin{equation*}
A^{i}=A^{i}(x,y),
\end{equation*}
and 
\begin{equation*}
A_{i}=A_{i}(x,y).
\end{equation*}

Variational procedures applied to the above naturally lead to additional
terms in Lorentz force

\begin{eqnarray*}
\dfrac{dy^{i}}{dt}+\Gamma _{~jk}^{i}y^{j}y^{k} &=&\dfrac{q}{c}%
F_{~h}^{i}y^{h}+\dfrac{q}{c}\tilde{F}_{~j}^{i}\dfrac{dy^{j}}{dt},~\ \ y^{i}=%
\dot{x}^{i} \\
F_{~h}^{i} &=&g^{ij}F_{jh},~\ \ F_{jh}=\dfrac{\partial A_{h}}{\partial x^{j}}%
-\dfrac{\partial A_{j}}{\partial x^{h}}, \\
\tilde{F}_{~j}^{i} &=&g^{ih}\tilde{F}_{hj}~\ \tilde{F}_{hj}=-\dfrac{\partial
A_{h}}{\partial y^{j}}
\end{eqnarray*}
as well as the appearance of a correction to the usual expression of
currents: 
\begin{equation*}
D_{\tfrac{\partial }{\partial x^{i}}}F^{ki}+D_{\tfrac{\partial }{\partial
y^{a}}}F^{ka}=J^{k}.
\end{equation*}

In the present paper, we investigate the case of Finslerian spaces whose
metrics are obtained by a small (linearly approximable) deformation of
metric tensors whose components do not depend on positional variables (%
\textit{locally Minkowskian metrics}) and coordinate changes which preserve
the positional independence of the undeformed metric. The construction will
be generalized to arbitrary Finsler spaces in future works.

\bigskip

\section{Weak Finslerian deformation of locally Minkowskian metrics}

A \textit{locally Minkowskian }Finsler space is a Finsler space $(M,\mathcal{%
F})$ with the property that there exists a local coordinate system with
respect to which the components of the corresponding metric tensor do not
depend on positional variables, but only on directional ones: 
\begin{equation*}
\gamma _{ij}=\gamma _{ij}(y).
\end{equation*}

In the following, we shall only consider coordinate changes which preserve
this property.

One of the properties of locally Minkowski spaces is projective flatness,
namely, their geodesics are straight lines:

This type of metrics includes as particular cases:

\begin{itemize}
\item Minkowski metric $\gamma =diag(1,-1,-1,-1);$

\item Berwald-Moor 4-dimensional metric, \cite{Pavlov}, \cite{Leb}, \cite{BB}%
, \cite{BBL}.
\end{itemize}

\bigskip

Let us consider the space $\mathbb{R}^{4},$ endowed with linear coordinate
changes. Let $(x,y)=(x^{i},y^{a})_{i,a=\overline{1,4}},$ $y^{i}=\dfrac{%
\partial x^{i}}{\partial t}$ ($t$ is a parameter), $i=1,..,4$ be the
coordinates in a local frame of $T\mathbb{R}^{4}\equiv \mathbb{R}^{8}$.

Let $g$ be a small (linearly approximable)\ deformation of a locally
Minkowskian metric: 
\begin{equation}
g_{ij}(x,y)=\gamma _{ij}(y)+\varepsilon _{ij}(x,y).  \label{metric}
\end{equation}

We suppose that this metric tensor is Finslerian in the sense of \cite%
{Rashewski}, this is, 
\begin{equation*}
g_{ij}=\dfrac 12\dfrac{\partial ^2\mathcal{F}^2}{\partial y^i\partial y^j}
\end{equation*}
for some 2-homogeneous in $y$ function $\mathcal{F}=\mathcal{F}(x,y),$ and $%
g_{ij}$\ is nondegenerate. We denote by $_{,~k}$ (with commas) partial
derivation w.r.t. $x^k$ and with dots $_{\cdot a}$, partial derivation by
the directional variable $y^a.$ Whenever convenient - and just in order to
point out the difference, we will denote the indices corresponding to $y$
with $a,b,c,...$ and those corresponding to $x$ with $i,j,k,...,$ (though,
they run over the same set $\{1,2,3,4\}$). Let 
\begin{equation*}
\Gamma _{~jk}^i=\dfrac 12g^{ih}(g_{hj,k}+g_{hk,j}-g_{jk,h})
\end{equation*}
denote the usual Christoffel symbols (with respect to $x$) of $g.$

In our case, $\Gamma _{~jk}^{i}$ depend on both $x$ and $y:$%
\begin{equation*}
\Gamma _{~jk}^{i}=\Gamma _{~jk}^{i}(x,y).
\end{equation*}

Let $_{|k}$ denote covariant derivation with respect to $x^{k}$ $:$%
\begin{equation}
X_{~~|k}^{i}=X_{~~,k}^{i}+\Gamma _{~jk}^{i}X^{j}.  \label{covar-deriv}
\end{equation}

In the following, we shall also need the \textbf{Cartan tensor} $C_{ijk},$ 
\cite{Shen}: 
\begin{equation}
C_{ijk}=\dfrac{1}{2}(g_{ij\cdot k}+g_{ik\cdot j}-g_{jk\cdot i})=\dfrac{1}{2}%
g_{ij\cdot k}.  \label{Cartan}
\end{equation}

Also, let 
\begin{equation*}
X^{i}|_{a}=X_{~\cdot a}^{i}+C_{~ja}^{i}X^{j}
\end{equation*}
denote covariant derivative with respect to $y^{a}.$

\section{\label{Lorentz_force}Lorentz force}

\subsection{Variational principle}

The equations of electrodynamics can be obtained from the variational
procedure applied to a Lagrangian. In \textbf{isotropic spaces}, the
Lagrangian is, \cite{MB}, 
\begin{equation*}
L(x,y)=\dfrac{1}{2}g_{ij}(x)y^{i}y^{j}+\dfrac{q}{c}A_{i}(x)y^{i},~\ \ y^{i}=%
\dot{x}^{i}.
\end{equation*}
where $q$\ is the electric charge, and $A_{i}(x)$ are the covariant
components of the 4-vector potential.

\bigskip

In order to obtain Lorentz force in \textbf{Finslerian spaces}, let us
consider the Lagrangian

\begin{equation*}
L=L_{0}+\dfrac{q}{c}L_{1},
\end{equation*}
where 
\begin{equation*}
L_{0}=\dfrac{1}{2}g_{ij}(x,y)y^{i}y^{j}
\end{equation*}
($g_{ij}$ can be chosen as a general Finslerian metric tensor) and $%
L_{1}=L_{1}(x,y)$ is a scalar function which is 1-homogeneous in the
directional variables: $L_{1}(x,\lambda y)=\lambda L_{1}(x,y),~\forall
\lambda \in \mathbb{R}$. Let 
\begin{equation*}
A_{i}(x,y):=\dfrac{\partial L_{1}}{\partial y^{i}}.
\end{equation*}
Then: 
\begin{equation*}
L_{1}=A_{j}(x,y)y^{j}
\end{equation*}
and our Lagrangian is written as

\begin{equation}
L(x,y)=\dfrac 12g_{ij}(x,y)y^iy^j+\dfrac qcA_i(x,y)y^i,~\ y^i=\dot x^i,
\label{L}
\end{equation}
where $A_j=A_j(x,y)$ is now a \textbf{direction dependent} potential.

The components of the covector field $A_i=A_i(x,y)$ are 0-homogeneous
functions in $y,$ and possess the property 
\begin{equation}
A_{i\cdot k}y^i=0.
\end{equation}

\bigskip

The Euler-Lagrange equations 
\begin{equation*}
\dfrac{\partial L}{\partial x^{i}}-\dfrac{d}{dt}(\dfrac{\partial L}{\partial
y^{i}})=0
\end{equation*}
attached to $L$ lead to

\begin{equation}
g_{kh}\left( \dfrac{dy^{h}}{dt}+\Gamma _{~jl}^{h}y^{j}y^{l}\right) +\dfrac{q%
}{c}(A_{k,h}-A_{h,k})y^{h}+\dfrac{q}{c}A_{k\cdot h}\dfrac{dy^{h}}{dt}=0.
\label{Lorentz-rough}
\end{equation}

Let: 
\begin{equation}
F_{kh}=A_{h,k}-A_{k,h}  \label{classical-F}
\end{equation}

We have thus obtained

\begin{proposition}
The extremal curves $t\mapsto (x^i(t)):[0,1]\rightarrow \mathbb{R}^4$ of the
Lagrangian (\ref{L}) are given by 
\begin{equation}
\dfrac{dy^i}{dt}+\Gamma _{~jk}^iy^jy^k=\dfrac qcF_{~h}^iy^h-\dfrac
qcg^{ik}A_{k\cdot h}\dfrac{dy^h}{dt},  \label{Lorentz1}
\end{equation}
\end{proposition}

\begin{remark}
The term $F^i(x,y)\equiv \dfrac qcg^{ik}F_{kh}y^h$ is present also in the
isotropic case (see \cite{MB}). But the last one, 
\begin{equation*}
\tilde F^i(x,y):=-\dfrac qcg^{ik}(x,y)A_{k\cdot h}(x,y)\dfrac{dy^h}{dt}
\end{equation*}
can only appear in anisotropic ones.
\end{remark}

The usual interpretation of the extremal curves is the equation of motion.
Therefore, the expression in the rhs of (\ref{Lorentz1}) presents the
Lorentz force in the anisotropic space. We see that its first term which is
common with the isotropic case is proportional to velocity, while the second
term is proportional to acceleration which brings to mind the idea of an
\textquotedblright inertial\ force\textquotedblright\ in the accelerated
reference frame.

Let us designate 
\begin{equation}
\tilde{F}_{ia}:=-A_{i\cdot a},~\ \ \tilde{F}_{ai}=A_{i\cdot a},
\label{mixed-F}
\end{equation}
where we denote by $a,b,c,d,...$ indices corresponding to derivation by
directional variables.

Then $\tilde{F}_{ia}$\ is (-1) homogeneous in the directional variables: 
\begin{equation*}
\tilde{F}_{ia}(x,\lambda y)=\dfrac{1}{\lambda }\tilde{F}_{ia}(x,y),~\ \ \ \
\lambda \in \mathbb{R}.
\end{equation*}

Then the relation between $\tilde{F}_{ia}\emph{\ }$and\ the new term in (\ref%
{Lorentz1}) is 
\begin{equation*}
\tilde{F}^{i}=\dfrac{q}{c}\tilde{F}_{~a}^{i~}\dfrac{dy^{a}}{dt},
\end{equation*}

and we have thus obtained an antisymmetric 2-form on $T\mathbb{R}^{4}:$%
\begin{equation}
F=F_{ij}dx^{i}\wedge dx^{j}+\tilde{F}_{ia}dx^{i}\wedge dy^{a}.
\label{2-form}
\end{equation}

The above is nothing but the exterior derivative of the 1-form $%
A=A_{i}(x,y)dx^{i}+0\cdot dy^{a}$ on $T\mathbb{R}^{4}:$%
\begin{equation}
F=dA.  \label{good-looking1}
\end{equation}

\bigskip

\textbf{Conclusion: }Direction dependent electromagnetic potentials lead in
a natural way to a correction to the expression of the electromagnetic
tensor.

\bigskip

\begin{example}
\begin{enumerate}
\item In the particular case when the \textit{covariant} components of $A$
do not depend on direction, 
\begin{equation*}
A_i=A_i(x),
\end{equation*}
then $\tilde F_{ia}=0$ and we get the regular expression of Lorentz force.

\item A simple, but nontrivial particular case is obtained when the \textit{%
contravariant} components of the potential 4-vector do not depend on the
directional variables: 
\begin{equation*}
A^i=A^i(x),
\end{equation*}
taking into account the $y$-dependence of the perturbed metric tensor $%
g_{ij},$ we get that the covariant components of $A$ are direction
dependent: 
\begin{equation*}
A_i=g_{ij}(x,y)A^j\Rightarrow ~A_i=A_i(x,y).
\end{equation*}
\end{enumerate}
\end{example}

The new term to appear in Lorentz force is then 
\begin{equation*}
\tilde{F}^{i}=\dfrac{q}{c}\tilde{F}_{~a}^{i~}\dfrac{dy^{a}}{dt}=-\dfrac{q}{c}%
g^{ih}A_{h\cdot a}\dfrac{dy^{a}}{dt},
\end{equation*}
which leads to 
\begin{equation}
\tilde{F}^{i}=-2\dfrac{q}{c}C_{~ja}^{i}A^{j}\dfrac{dy^{a}}{dt},
\label{exampleL1}
\end{equation}
and 
\begin{equation}
\tilde{F}_{ia}=-2C_{ija}A^{j}.  \label{exampleL2}
\end{equation}

\begin{example}
In particular, if $\gamma =diag(1,-1,-1,-1)$ is the Minkowski metric, and 
\begin{equation*}
g_{ij}=\gamma _{ij}+\varepsilon _{ij}(x,y),
\end{equation*}
where\textrm{\ }$\varepsilon _{ij}(x,y)$ is a small Finslerian perturbation,
then the above is 
\begin{equation}
\tilde F_{ia}=-\varepsilon _{ij\cdot a}A^j,  \label{appl-f}
\end{equation}
\end{example}

hence its values are small. For other locally Minkowskian metrics, the new
term $\tilde F_{ia}$ is not necessarily small.

\begin{example}
For the case when:

\begin{itemize}
\item $\gamma _{ij}$ is the Berwald-Moor Finslerian metric $\gamma
_{ij}=\dfrac 12\dfrac{\partial \mathcal{F}^2}{\partial y^i\partial y^j},$ $%
\mathcal{F}=\sqrt[4]{y^1y^2y^3y^4}$

\item $\varepsilon _{ij}=0,$ that is, 
\begin{equation*}
g_{ij}=\dfrac 12\dfrac{\partial \mathcal{F}^2}{\partial y^i\partial y^j},
\end{equation*}
and

\item $A^i=A^i(x),$ then the correction $\tilde F_{ia}$ is 
\begin{equation*}
\tilde F_{ia}=-A_{i\cdot a}=-\dfrac \partial {\partial
y^a}(g_{il}A^l)=-2C_{ila}A^l,
\end{equation*}
where the Cartan tensor $C_{ila}=\dfrac 14\dfrac{\partial ^3\mathcal{F}^2}{%
\partial y^i\partial y^l\partial y^a}$ has the explicit values 
\begin{equation*}
C_{ila}=\alpha \dfrac{\mathcal{F}^2}{y^iy^ly^a},~\ \ \alpha =\left\{ 
\begin{array}{c}
\dfrac 3{32}~\ \ ~\ \ \ \ \ \ \ \ \ \ \ if~\ \ i=j=k \\ 
-\dfrac 1{32}~\ \ ~\ \ \ \ \ \ \ \ \ \ if~\ \ i=j\not =k \\ 
\dfrac 1{32}~\ \ if~\ \ ~\ \ \ \ i\not =j\not =k\not =i.%
\end{array}
\right.
\end{equation*}
\end{itemize}
\end{example}

Here we see that $C_{ila}$ are not necessarily small.

\subsection{New term - ''electromagnetic'' vs. ''metric''}

Let us now have a look at equation (\ref{Lorentz-rough}): 
\begin{equation*}
g_{kh}\dfrac{Dy^{h}}{dt}=\dfrac{q}{c}F_{kh}y^{h}-\dfrac{q}{c}A_{k\cdot h}%
\dfrac{dy^{h}}{dt},~\ y=\dot{x}.
\end{equation*}

From the mathematical point of view, we can interpret the last term in two
ways:

\begin{itemize}
\item Since it appears multiplied by the acceleration $\dfrac{dy^h}{dt}$ and
moreover, since $A_{k\cdot h}=A_{h\cdot k},$ we can \textquotedblright
stick\textquotedblright\ it to the metric: 
\begin{equation*}
(g_{kh}+\dfrac qcA_{k\cdot h})\dfrac{dy^h}{dt}+\Gamma _{khl}y^hy^l=\dfrac
qcF_{kh}y^h.
\end{equation*}

and get a new metric tensor 
\begin{equation}
\tilde g_{kh}=g_{kh}+\dfrac qcA_{k\cdot h}  \label{met-new}
\end{equation}

(if the matrix $(\tilde g_{kh})$ is invertible), with the property 
\begin{equation*}
\tilde g_{kh}y^ky^h=g_{kh}y^ky^h=\mathcal{F}^2
\end{equation*}
\end{itemize}

With this, we can write the equation of motion as 
\begin{equation}
\dfrac{Dy^{i}}{dt}=\tilde{g}^{ik}\dfrac{q}{c}F_{kh}y^{h}
\label{met-new-geod}
\end{equation}

and the obtained expression for Lorentz force $\tilde{g}^{ik}\dfrac{q}{c}%
F_{kh}y^{h}$ differs from the case of isotropic perturbation $\dfrac{q}{c}%
g^{ik}F_{kh}y^{h}$ due to the new metric (\ref{met-new}).

\begin{itemize}
\item Also, we might leave the metric as it is and move the third term in
the right hand side and interpret it as a new term added to Lorentz force: 
\begin{equation*}
g_{kh}\dfrac{Dy^h}{dt}=\dfrac qcF_{kh}y^h-\dfrac qcA_{k\cdot h}\dfrac{dy^h}{%
dt}.
\end{equation*}

This would yield 
\begin{equation}
\dfrac{dy^i}{dt}+\Gamma _{~jk}^iy^jy^k=\dfrac qc(F_{~h}^{i~}y^h+\tilde
F_{~a}^{i~}\dfrac{dy^a}{dt}).  \label{Lorentz-weak}
\end{equation}
\end{itemize}

with the influence of the anisotropy given by the second term in the rhs.
Notice, that $\dfrac{q}{c}(F_{~h}^{i~}y^{h}+\tilde{F}_{~a}^{i~}\dfrac{dy^{a}%
}{dt})$ is equal to $\dfrac{q}{c}g^{ik}(F_{kh}y^{h}-A_{k\cdot a}\dfrac{dy^{a}%
}{dt})$\ given by eq.(\ref{Lorentz1}) since $F_{~h}^{i~}=g^{ik}F_{kh},$\ $%
\tilde{F}_{~a}^{i~}=g^{ik}A_{k\cdot a}$. In eq.(\ref{met-new-geod}) the term 
$\dfrac{q}{c}g^{ik}A_{k\cdot h}\ddot{x}^{h}$\ was brought to the left hand
side of the equation of motion and \textquotedblright
swallowed\textquotedblright\ into the metric - the new \textquotedblright
metric\textquotedblright\ was denoted by \thinspace $\tilde{g}_{ik}$. This
illustrates the remark concerning the equivalence principle made in \cite%
{Newt-7}, applied to a (curved) space with an electromagnetic field.

\section{Homogeneous Maxwell equations}

Let us consider again the 2-form (\ref{2-form}) 
\begin{equation*}
F(x,y)=dA(x,y)=\dfrac{1}{2}F_{ij}(x,y)dx^{i}\wedge dx^{j}+\tilde{F}%
_{ia}(x,y)dx^{i}\wedge dy^{a}.
\end{equation*}

We immediately get:

\begin{proposition}
There holds
\end{proposition}

\begin{equation}
F_{ij,k}+F_{ki,j}+F_{jk,i}=0;  \label{Max-1}
\end{equation}

which is just the homogeneous Maxwell equation or, in terms of covariant
derivatives (\ref{covar-deriv} ), 
\begin{equation*}
F_{ij|k}+F_{ki|j}+F_{jk|i}=0.
\end{equation*}

There also hold the equalities 
\begin{eqnarray}
\tilde{F}_{ia,k}+\tilde{F}_{ki\cdot a}+\tilde{F}_{ak,i} &=&0;  \notag \\
\tilde{F}_{ia\cdot b}+\tilde{F}_{bi\cdot a} &=&0,  \notag
\end{eqnarray}
where $F_{ia\cdot b}=\dfrac{\partial F_{ia}}{\partial y^{b}}.$ The above two
relations, together with (\ref{Max-1}) mean actually that the exterior
derivative of $F$ is 0: 
\begin{equation}
dF=0.  \label{dF}
\end{equation}

\bigskip

\section{Currents in anisotropic spaces}

In the classical Riemannian case, the inhomogeneous Maxwell equations can be
obtained by means of the variational principle applied to

\begin{equation*}
\int (\alpha F_{ij}F^{ij}-\beta j^kA_k)\sqrt{-g}d\Omega ,
\end{equation*}
\cite{RBS}, where $\alpha $ and $\beta $ are constants, $g=\det (g_{ij})$
and $\Omega =dx^1dx^2dx^3dx^4.$ Taking into account that in our case, at
least one of the quantities $F_{ij},$ $F^{ij}$ depends on $y,$ the whole
integrand depends on $y,$ and is actually defined on some domain in $\mathbb{%
R}^8.$

Let 
\begin{equation*}
u^{a}=\dfrac{1}{H}y^{a},
\end{equation*}
where $H$ is a constant, $[H]=\dfrac{1}{\sec },$ meant to "adjust"
measurement units as to have $[x^{i}]=[u^{a}],$ hence also $[F_{ij}]=[\tilde{%
F}_{ia}].$ So, let us consider $A_{i}=A_{i}(x,u)$ and the integral of action 
\begin{equation*}
I=\int (\alpha F_{\lambda \mu }F^{\lambda \mu }-\beta j^{k}A_{k})\sqrt{G}%
d\Omega ,
\end{equation*}

where $\lambda ,\mu \in \{i,j,a,b\},$ $G=\det (G_{\alpha \beta })$ is the
Sasaki lift of $g$ to $T\mathbb{R}^{4}\equiv \mathbb{R}^{8},$ \cite{Shen}: 
\begin{equation*}
G_{\alpha \beta }(x,u)=g_{ij}(x,u)dx^{i}\otimes
dx^{j}+g_{ab}(x,u)du^{a}\otimes du^{b}
\end{equation*}

and $\Omega =~\underset{i,a}{\Pi }dx^{i}du^{a}$ gives the volume form on $%
\mathbb{R}^{8}.$

\textit{Remark:} The product $F_{\lambda \mu }F^{\lambda \mu }$ is in our
case 
\begin{equation*}
F_{\lambda \mu }F^{\lambda \mu }=F_{ij}F^{ij}+\tilde{F}_{ia}\tilde{F}^{ia}.
\end{equation*}

Taking variations with respect to $A_k$ in the above, we get, in terms of
covariant derivatives with respect to the metrical connection $D\Gamma
=(\Gamma _{~jk}^i,C_{~jk}^i),$%
\begin{equation*}
F_{~~~|i}^{ki}+\tilde F^{ka}|_a=J^i,
\end{equation*}
where \ $_{|i}=D_{\tfrac \partial {\partial x^i}},$ $|_a~=D_{\tfrac \partial
{\partial y^a}}.$

\textbf{Conclusion: }In comparison to the case of isotropic spaces, there
appears a new term in the expression for the current, namely, 
\begin{equation}
\zeta ^k=\tilde F^{ka}|_a.  \label{new-cur}
\end{equation}

This means that in an anisotropic space the measured fields would correspond
to an efective current consisting of two terms: one is the current provided
by the experimental environment, the other is the current corresponding to
the anisotropy of space.

\bigskip

\textbf{Examples:}

\begin{enumerate}
\item If $A_i=A_i(x),$ then we get $\tilde F_{ia}=0$ and 
\begin{equation*}
\zeta ^k=0,~\ \ F_{~~~|i}^{ki}=J^k.
\end{equation*}

\item Already a nontrivial example is obtained if 
\begin{equation*}
A^i=A^i(x);
\end{equation*}
then we have shown above that 
\begin{equation*}
\tilde F_{ia}=-2C_{ija}A^j.
\end{equation*}
\end{enumerate}

Then, $\tilde{F}^{ia}=-2C_{~~j}^{ia}A^{j}$ and 
\begin{equation*}
\zeta
^{k}=-2(C_{~~j}^{ka}A^{j})|_{a}=-2C_{~~~j}^{ka}|_{a}A^{j}-2C_{~~~j}^{ka}C_{~ha}^{j}A^{h}.
\end{equation*}

\bigskip

The presence of the last current in the experimental situation could be
noticed if $\left| \tilde F^{ka}|_a\right| \approx \left|
F_{~~~|i}^{ki}\right| $.

In the case of \textit{deformed Minkowski metric}, we get that 
\begin{equation}
\zeta ^k=-2\varepsilon _{~~~~\cdot ja}^{ka}A^j.  \label{appl-c}
\end{equation}

Consequently, the current $\tilde F^{ia}$ can be noticed when the
\textquotedblright regular\textquotedblright\ current is small and $\left|
2\varepsilon _{~~~~\cdot ja}^{ka}A^j\right| \approx \left|
F_{~~~|i}^{ki}\right| .$

If we take instead a deformed Berwald-Moor metric, then the Cartan tensor
components are no longer small, and hence the correction $\zeta ^{k}$ is
generally not small, and it could be noticed even if the regular current is
not small.

\bigskip

\ \textbf{Conclusion: }An anisotropic space with electromagnetic field
possesses inherent currents that could produce observable fields.

\bigskip

\textbf{Comparison to existent results:}

R. Miron and collaborators, \cite{Lagrange} defined electromagnetic tensors
in Lagrange spaces formally, by means of nonlinear (and linear) connections
on the tangent bundle $TM:$

\begin{equation*}
F_{ij}=\dfrac 12(y_{j|i}-y_{i|j}),f_{ab}=\dfrac 12(y_b|_a-y_a|_b)
\end{equation*}
where $N_{~j}^i,$ $L_{~jk}^i,~C_{~bc}^a$ are the coefficients of the Kern
nonlinear connection and respectively, the canonical metrical linear
connection, \cite{Lagrange}, on $TM$ (in the case of Finsler spaces, $%
(N_{~j}^i,$ $L_{~jk}^i,~C_{~bc}^a)$ provide the Cartan connection, \cite%
{Shen}).

We notice\textrm{\ }there the appearance of new quantities $f_{ab}$ (and
additional Maxwell equations) in comparison to the Riemannian case: 
\begin{eqnarray*}
F_{ji|k}+F_{kj|i}+F_{ik|j} &=&0,~\ \ f_{ab|c}+f_{ca|b}+f_{bc|a}=0, \\
F_{~~~|j}^{ij} &=&J^{i},~\ \ f^{ab}|_{b}=j^{a}.
\end{eqnarray*}%
Also, in the cited work is investigated the case of an electromagnetic
tensor arising from a potential 
\begin{equation*}
A_{i}=A_{i}(x)
\end{equation*}%
depending just on the positional variables $x^{i}$. In this case, the
authors show that the resulting Maxwell equations and the resulting
expression for Lorentz force are formally identical to the usual ones in
Riemannian spaces - no additional terms appear.

\bigskip

In our approach, we consider \textbf{direction dependent potentials}. The
components and the corrections due to anisotropy to the electromagnetic
tensor appear from variational approaches. The new appearing terms are to be 
\textit{added} to the currents, not regarded as separate quantities: 
\begin{equation*}
F_{~~~|j}^{ij}+\tilde{F}_{~~~\cdot a}^{ia}=J^{i}.
\end{equation*}

\bigskip

We could also relate our components of the electromagnetic tensor to linear
connections. Namely, let $D\Gamma =(N_{~i}^{a},L_{~jk}^{i},C_{~jk}^{i})$
denote the Cartan connection, \cite{Shen}, \cite{Lagrange}, determined by
the Finslerian function $g_{ij};$ let us define the following tensor fields: 
\begin{equation*}
X_{~jk}^{i}=\dfrac{2q}{c}\dfrac{g^{ih}}{\mathcal{F}}\dfrac{\partial }{%
\partial y^{k}}(\mathcal{F}A_{h|j}),
\end{equation*}
where the $x$-covariant derivative $A_{h|j}$ is taken with respect to the
Cartan connection of $g:$ $A_{h|j}=\dfrac{\delta A_{h}}{\delta x^{j}}%
-L_{~hj}^{l}A_{l}.$ Also, let 
\begin{equation*}
\Omega _{rj}^{ih}=\dfrac{1}{2}(\delta _{r}^{i}\delta _{j}^{h}-g_{rj}g^{ih})
\end{equation*}
denote the Obata operators, \cite{Lagrange}, of $g.$

Let us fix the nonlinear connection $N$ as the Cartan one and define the
linear connection $\tilde{D}\Gamma (N)=(\tilde{L}_{~jk}^{i},\tilde{C}%
_{~jk}^{i})$ by 
\begin{equation*}
\tilde{L}_{~jk}^{i}=\Gamma _{~jk}^{i}+\Omega _{rj}^{ih}X_{~hk}^{r},~\ \ 
\tilde{C}_{~jk}^{i}=C_{~jk}^{i}-\dfrac{q}{c}g^{ih}A_{h\cdot jk}.
\end{equation*}

We denote $x$- and $y$- covariant derivatives with respect to this
connection with $_{||k}$ and $||_{k}$ respectively.

Then, by direct computation, one can check the following properties:

\begin{proposition}
\begin{enumerate}
\item A curve $t\mapsto (x^i(t),\dot x^i(t))$ on $T\mathbb{R}^4$ is an
autoparallel curve of $\tilde D\Gamma (N)$ if and only if its projection $%
t\mapsto x^i(t)$ on the base manifold $\mathbb{R}^4$ is a solution of
Lorentz equation (\ref{Lorentz1}).

\item $\tilde D\Gamma (N)$ is h-metrical: $g_{ij||k}=0;$

\item In the adapted basis $(\dfrac \delta {\delta x^i},\dfrac \partial
{\partial y^k})$, the local coordinates of the electromagnetic tensor are
given by: 
\begin{eqnarray*}
y_{i||j}=\dfrac{-1}2(y_{j||i}-y_{i||j})=-\dfrac
qc(A_{j||i}-A_{i||j})=:-\dfrac qc\mathbf{F}_{ij}. \\
y_i||_k=g_{ik}-\dfrac qcA_{i\cdot k},
\end{eqnarray*}
\newline
\end{enumerate}
\end{proposition}

\section{The Berwald-Moor case}

In the following, we shall develop an approach for obtaining a
generalization of the expression for the Lorentz force in the case of the
Berwald-Moor quartic Finslerian function $\mathcal{F=}\sqrt[4]{y^1y^2y^3y^4}$%
\ , \cite{Pavlov}, \cite{Leb}, \cite{BB}, \cite{BBL}, \cite{projective}, in
terms of the 4-scalar product introduced in \cite{Pavlov}.

The 4-scalar product 
\begin{equation*}
<U,V,W,Y>~=~G_{ijkl}U^iV^jW^kY^l,~\ \ \ \ \forall U,V,W,Y\in \mathcal{X}(%
\mathbb{R}^4),
\end{equation*}
where the components $G_{ijkl}$ are, \cite{Pavlov}, 
\begin{equation*}
G_{ijkl}=\left\{ 
\begin{array}{l}
\dfrac 1{4!}~\ \text{\ if }i,j,k,l\text{ are all different from each other}
\\ 
0,~\ \ \ \ \ \text{\ \ \ \ \ \ \ \ \ \ \ \ \ \ elsewhere}%
\end{array}
\right.
\end{equation*}

induces a (direction dependent) pseudo-scalar product

\bigskip

if we specify two of the 4 vectors involved.\ For instance, for $W=Y=y,$we
get 
\begin{equation*}
<U,V>~:=~<U,V,y,y>~=h_{ij}U^iV^j,
\end{equation*}
where the \textit{flag (polynomial) metric tensor }$h=h(y)$ has the local
components 
\begin{equation*}
h_{ij}=G_{ijkl}y^ky^l=:G_{ij00},
\end{equation*}
equivalently, \cite{projective}, 
\begin{equation*}
h_{ij}=\dfrac 1{12}\dfrac{\partial ^2\mathcal{F}^4}{\partial y^i\partial y^j}%
.
\end{equation*}

We notice that the components $h_{ij}$ are 2-homogeneous ploynomials in the
directional variables $y^{i}$ and that 
\begin{equation*}
h_{ij}y^{i}y^{j}=\mathcal{F}^{4}
\end{equation*}

\bigskip

The Lagrangian in Section \ref{Lorentz_force}, providing Lorentz force can
be written as 
\begin{equation*}
L=~\dfrac{1}{2}<y,y>+\dfrac{q}{c}<A,y>.
\end{equation*}

If we use in the above, instead of the classical Finslerian metric tensor $%
g_{ij}$, the polynomial metric $h_{ij},$ then we get a specific
(generalized) Lagrangian for the Berwald-Moor case: 
\begin{equation*}
L=\dfrac 12h_{ij}y^iy^j+\dfrac qch_{ij}A^iy^j.
\end{equation*}

\textbf{Remark: }The first term in the above is $\dfrac{1}{2}%
h_{ij}y^{i}y^{j}=\dfrac{1}{2}\mathcal{F}^{4}.$

\bigskip

The second term is 
\begin{equation*}
\dfrac{q}{c}h_{ij}A^{i}y^{j}=\dfrac{q}{c}G_{ijkl}A^{i}y^{j}y^{k}y^{l}.
\end{equation*}

\bigskip

Let us denote: 
\begin{equation*}
A_{jkl}=G_{ijkl}A^{i}.
\end{equation*}
The above defined tensor is totally symmetric. Its nonvanishing components
are: 
\begin{equation*}
A_{123}=\dfrac{1}{4!}A^{4},~\ \ A_{124}=\dfrac{1}{4!}A^{3},~\ A_{134}=\dfrac{%
1}{4!}A^{2},~\ \ A_{234}=\dfrac{1}{4!}A^{1}.
\end{equation*}

Moreover, if $A^{i}=A^{i}(x)$ are only position dependent, then $A_{jkl}$
depend only on $x$ and conversely. In the following, we shall assume this,
namely 
\begin{equation*}
A_{jkl}=A_{jkl}(x)
\end{equation*}

We get, thus, the following Lagrangian in case of the Berwald-Moor function 
\begin{equation}
L=\dfrac{1}{2}\mathcal{F}^{4}+\dfrac{q}{c}A_{ijk}(x)y^{i}y^{j}y^{k}.
\label{BM_Lagrangian}
\end{equation}

Let us obtain the terms of the Euler-Lagrange equation 
\begin{eqnarray*}
&&\dfrac{\partial L}{\partial x^m}=\dfrac qcA_{ljk,m}y^ly^jy^k; \\
&&\dfrac{\partial L}{\partial y^m}=\dfrac 12\dfrac{\partial \mathcal{F}^4}{%
\partial y^m}+3\dfrac qcA_{mjk}(x)y^jy^k~\Rightarrow ~ \\
&&\dfrac d{dt}(\dfrac{\partial L}{\partial y^m})=\dfrac d{dt}(\dfrac 12%
\dfrac{\partial \mathcal{F}^4}{\partial y^m})+3\dfrac
qcA_{mjk,l}y^ly^jy^k+\dfrac qc6A_{jmk}\dfrac{dy^j}{dt}y^k= \\
&=&(\dfrac{12}2h_{mj}\dfrac{dy^j}{dt}+6\dfrac qcA_{jmk}y^k)\dfrac{dy^j}{dt}%
+\dfrac qc(A_{mjk,l}+A_{mlk,j}+A_{mjl,k})y^ly^jy^k.
\end{eqnarray*}

Therefore, we get

\begin{proposition}
The extremal curves for the Lagrangian (\ref{BM_Lagrangian}) are given by 
\begin{equation*}
6(h_{mj}+\dfrac qcA_{mj0})\dfrac{dy^j}{dt}+\dfrac
qc(A_{mjk,l}+A_{mlk,j}+A_{mjl,k}-A_{ljk,m})y^ly^jy^k=0,
\end{equation*}
where $A_{mj0}=A_{mjk}y^k.$
\end{proposition}

\bigskip

Let us denote 
\begin{equation}
F_{mjkl}=\dfrac{1}{6}(A_{mjk,l}+A_{mlk,j}+A_{mjl,k}-A_{ljk,m}).
\label{4-index_F}
\end{equation}
The above relation defines a 4-covariant tensor field, which is symmetric in
its 3 last indices: $F_{mjkl}=F_{mkjl}=F_{mklj}$ etc.

By contracting it twice with $y^{k},y^{l},$ we get an antisymmetric tensor: 
\begin{equation*}
F_{mj00}:=F_{mjkl}y^{k}y^{l}\Rightarrow ~F_{mj00}=F_{jm00}.
\end{equation*}

With the notation (\ref{4-index_F}), the equations of extremal curves can be
written as: 
\begin{equation*}
h_{mj}\dfrac{dy^j}{dt}+\dfrac qcF_{mjkl}y^jy^ky^l+\dfrac 16\dfrac qcA_{mj0}%
\dfrac{dy^j}{dt}=0.
\end{equation*}
If we raise indices by means of $h^{ij},$ the above is equivalent to : 
\begin{equation}
\dfrac{dy^i}{dt}+\dfrac qcF_{~jkl}^iy^jy^ky^l+\dfrac 16\dfrac qcA_{~j0}^i%
\dfrac{dy^j}{dt}=0.  \label{BM-geod}
\end{equation}

With this we can return to \ref{Lorentz1} and try to derive all the results
obtained in what follows there in the similar manner.

\bigskip

\section{Discussion}

The relativity principle states that there are no means to distinguish
between the inertial frames, but the equivalence principle goes even
further. Since we can't distinguish between an accelerated frame and such a
physical field as gravity, there is no reason to think that gravity is
velocity independent. If the acceleration is not rectilinear, the inertial
force acting on a body depends on the velocity of the body (e.g. Coriolis
force). Therefore, measuring gravity we can't be sure to which extent we
have accounted for the observable kinematics. From the mathematical point of
view it means that the space is anisotropic. Locally, it could mean the
existence of a preferable direction (e.g. the axis of a rotating reference
frame), but generally it demands the direction dependent metric. This leads
to the anisotropic geometrodynamics \cite{Newt-7} which appears to be able
to explain some well known paradoxes observed on the cosmological scale.

In this paper the ideas formulated in \cite{Newt-7} were developed for the
case when an additional physical (electromagnetic) field is present in such
an anisotropic space. Obviously, the notions of Lorentz force and electric
current in an accelerated frame (in an anisotropic space) should be
redefined with regard to the mentioned circumstances. This is performed for
the case of the linearly approximable anisotropic perturbation of the
locally Minkowski metric. We have found the expressions for the additional
terms both in the Lorentz force \ref{Lorentz1} and in the current \ref%
{new-cur}. It should be mentioned that it is important in which way -
covariant or contravariant - do the measurable (physical) variables
transform with the coordinates transformation. The calculations for the case
when the unperturbed locally Minkowski metric is Minkowski one lead to the
concrete expressions \ref{appl-f} and \ref{appl-c} that can be used in
observations.

Alongside with the weak anisotropic perturbation of the Minkowski metric we
regarded also the anisotropic Berwald-Moor metric as it was done in \cite%
{S-B}. Working in terms of the corresponding 4-scalar product, the geodesic
equations have obtained a new form and what could be called a
\textquotedblright Lorentz force\textquotedblright\ has also transformed.
This is due to the fact that the algebra chosen for the possible
interpretation of experimental measurements essentially differs from the
usual one. It is hard to say at once whether it is convenient or not and
whether it has any new perspectives because of the lack of the corresponding
language in the interpretation of physical notions.

\end{document}